\documentclass[onecolumn, notitlepage, superscriptaddress]{revtex4-1}

\usepackage{amsfonts}
\usepackage{amsmath,amssymb}
\usepackage{bm}
\usepackage{graphicx}
\usepackage{color}
\usepackage[colorlinks, urlcolor=blue, citecolor=blue, linkcolor=blue, pdfstartview=FitH]{hyperref}

\def\affiUK{School of Physics and Astronomy, University of Southampton, SO17 1BJ, Southampton, United Kingdom}
\def\affiSOLAB{Spin Optics Laboratory, St.-Petersburg State University, 198504,  Peterhof, St.-Petersburg, Russia}
\def\affiCNR{CNR-SPIN, Tor Vergata, viale del Politechnico 1, I-00133 Rome, Italy}
\newcommand\affiRQC{Russian Quantum Center, Novaya 100, 143025 Skolkovo, Moscow Region, Russia}
\def\affiITMO{ITMO University, 197101, St.-Petersburg, Russia}

\begin{document}

\title{Bunching of numbers in a non-ideal roulette: the key to winning
strategies}

\author{A.\ V.\ Kavokin}
\affiliation{\affiUK}  
\affiliation{\affiSOLAB} 
\affiliation{\affiRQC}
\affiliation{\affiCNR}

\author{A.\ S.\ Sheremet}
\affiliation{\affiRQC}
\affiliation{\affiITMO} 

\author{M.\ Yu.\ Petrov}
\affiliation{\affiSOLAB}

\date{\today}

\begin{abstract}
Chances of a gambler are always lower than chances of a casino in the case
of an ideal, mathematically perfect roulette, if the capital of the gambler
is limited and the minimum and maximum allowed bets are limited by the
casino. However, a realistic roulette is not ideal: the probabilities of
realisation of different numbers slightly deviate. Describing this deviation
by a statistical distribution with a width $\delta$ we find a critical $\delta$ 
that equalizes chances of gambler and casino in the case of a
simple strategy of the game: the gambler always puts equal bets to the last
N numbers. For up-critical $\delta$ the expected return of the roulette
becomes positive. We show that the dramatic increase of gambler's chances is
a manifestation of bunching of numbers in a non-ideal roulette. We also
estimate the critical starting capital needed to ensure the low risk game
for an indefinite time.
\end{abstract}

\maketitle
\nopagebreak

\section*{Introduction}

Roulette is one the most famous games of hasard. It is equally famous as a
mathematical toy model. In a European roulette, which we mostly discuss in
this Letter, a croupier spins a wheel in one direction, then spins a ball in
the opposite direction around a tilted circular track running around the
circumference of the wheel. The ball eventually loses momentum and falls
onto the wheel and into one of 37 colored and numbered pockets on the wheel 
\cite{book}. It has been shown a long time ago that a casino has always
better chances than a gambler for the simple reason that the probability of
realisation of any number is $1/37$, while in the case of win the gambler
collects his/her bet times 36 (for a recent review see \cite{review}). In
the American roulette, the chances of a gambler are even lower as the wheel
has 38 sectors (numbers from 0 to 36 and the double zero sector), while
betting on a number a player only collects 36 times the bet in the case of
win. Multiple strategies have been invented over centuries to circumvent
this simple mathematics that necessarily gives better chances to a casino
than to a gambler \cite{strategy}. As an example, the popular \emph{%
martingale} strategy relies on the game of chances (e.g. red and black),
where the bet is doubled in the case of win. The probability to win is $%
18/37<1/2$ each time, while gamblers try to increase their chances by
doubling their bets after each loss and coming back to the initial bet after
each win. Clearly, on a long time scale this strategy may only work if the
capital of the gambler is unlimited and the casino imposes no limits to the
bets, that is never the case. Other strategies are based on an intuitive
feeling of gamblers that a roulette might have memory. Assuming this, they
believe that if e.g. \textquotedblleft red\textquotedblright\ came 10 times
in row, there is a higher probability for \textquotedblleft
black\textquotedblright\ to come on the 11th spin, as the system must tend
to equalize the number of black and red outcomes. As there is no any
rational reason for roulette to have memory, this strategy is doomed to fail
on a long time-scale. In a similar way, none of \textquotedblleft winning
strategies\textquotedblright\ may provide better chances to a gambler than
to a casino on a long time-scale (number of spins tending to infinity). If a
gambler has an initial capital $C$ his/her chances to double this capital
are lower than the chances to lose it by at least 1/37 independently of the
adapted strategy.

Still, there are professional roulette players who manage to collect steady
profits from their matter. They exploit deviations of all realistic
mechanical roulettes from the ideal. These deviations may be caused by the
geometry of the roulette, correlation between the speed of spinning and the
initial velocity of the ball, as well as by human factors: a croupier is not
a random number generator, after all. Some well-trained croupiers may want
sending balls to preselected sectors of the wheel on purpose of overplaing
gamblers who place their bets in opposite sectors. All these factors result
in a non-uniform distribution of probabilities for realisation of different
numbers.

\section*{Theoretical model}

\begin{figure}[tbp]
\includegraphics[width=.6\columnwidth]{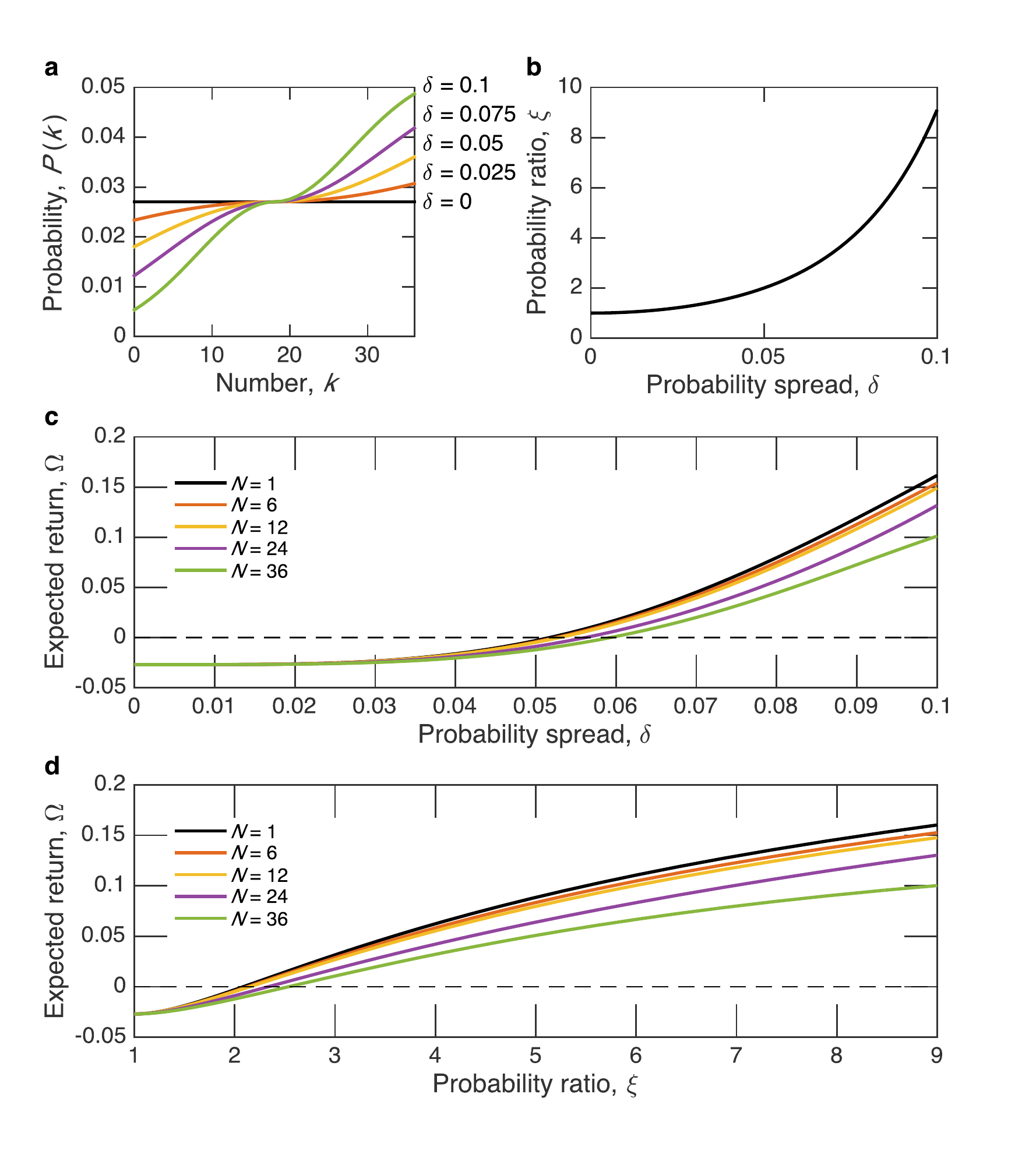}
\caption{(a) Probability distributions $P(k)$ calculated for different
values of the spread parameter $\protect\delta $. (b) Lowest-to-highest
probability ratio, $\protect\xi$, as a function of $\protect\delta$. (c,d)
Monte-Carlo simulation of $\Omega$ as a function of $\protect\delta$ (c) and 
$\protect\xi$ (d) calculated for different values of $N$, i.e. the length of
the sequence of the ``last'' numbers we bet on.}
\label{Fig1}
\end{figure}

Let us assume that the distribution of probabilities is a smooth function
with the maximum at 1/37 and Gaussian tails. Later we shall also consider a
linear distribution function for the sake of comparison. Renumbering all
numbers of the roulette in the order of increasing probability of
realisation, $k=0,1,\ldots ,36$ we set the probability for the number $k$ to
be realised as 
\begin{equation}
P\left( k\right) =%
\begin{cases}
\frac{1}{37}\exp \left( -\delta ^{2}\left( k-18\right) ^{2}/2\right) , & 
k\leq 18, \\ 
\frac{1}{37}\left[ 2-\exp \left( -\delta ^{2}\left( k-18\right)
^{2}/2\right) \right] , & k>18.%
\end{cases}
\label{eq_Gauss_distribution}
\end{equation}%
$k=0$ has the lowest probability with $P(0) =(1/37)\exp(-162\delta^{2})$, $%
k=36$ has the highest probability with $P(36) = 2/37 - P(0)$ and $P(18)=1/37$
for any $\delta$. In Fig.~\ref{Fig1}(a), we show the probability
distribution $P(k)$ for different values of $\delta$. It is instructive to
consider also the ratio between the highest and the lowest probability: 
\begin{equation}
\xi =\frac{P\left( 36\right) }{P\left( 0\right) }=2\exp \left( 162\delta
^{2}\right) -1,
\end{equation}%
which is shown in Fig.\ref{Fig1}(b). 
The following formalism is independent of the specific shape of $P(k)$,
while, of course, the numerical results may be affected by the choice of $%
P(k)$, as discussed below.

We underline that the correspondence between the numbers $k$ and the real
numbers on the roulette wheel is unknown to us, and we do not intend to know
it. Frequently, experienced gamblers try to collect statistics of
realisation of different numbers over several hundred spins, select
``happy'' numbers and then keep steadily playing on these numbers. This is a
time consuming method that sometimes meets disapproval by casino
administrations. Moreover, test measurements for ``happy numbers'' need to
be re-done frequently, as e.g. croupier gets replaced, speed of spinning may
be reset etc. A much simpler method suitable for an amateur consists in
betting on the last sequence of numbers realised. Many casinos display the
last numbers for the convenience of gamblers. We shall consider the case of
a gambler with a sufficient starting capital always playing the last $N$
numbers placing equal bets on each number. The advantage of this method is
that it does not rely on any statistics collected previously and easily
adapts to the changes imposed by the casino.

Here we prove that the method may be also considered as a convenient tool
for the measurement of the parameters $\delta$ and $\xi$\ characterising the
distribution of probabilities. In particular, we shall be looking for the
critical value of $\delta$ or $\xi$\ that provides a sufficient bunching of
numbers which equalizes the chances of gambler and casino.

Let us assume that the latest $N$ numbers chosen by a roulette are $%
k_{1},k_{2},...,k_{N}$. We will call this sequence $n$th realisation. We
note that some of these numbers may coincide. The probability of $n$th
realisation is given by $\prod\limits_{i=1}^{N}P(k_{i})$. We shall assume
that there are $j_{n}$ different numbers among the last $N$ numbers. $j_{n}$
varies from 1 to $N$. The probability to have on the $(N+1)$st position one
of the numbers $k_{i}$ $(i=1,2,...,N)$ is given by $\sum_{i=1}^{j_{n}}P(%
\overline{k}_{i})$. Here, $\overline{k}_{i}$ denote those $j_{n}$ numbers
that are different among $N$ last numbers. Hence, the probability to have
first the sequence $k_{1},k_{2},...,k_{N}$ and then one of the numbers $k_{i}
$, $(i=1,2,...,N)$ is given by 
\begin{equation}
\Xi_{n}=\prod\limits_{i=1}^{N}P(k_{i})\sum_{i=1}^{j_{n}}P(\overline{k}_{i}).
\end{equation}

Now, in order to find the probability of this event, we need to sum $\Xi_{n} 
$ over all possible sequences. This yields: 
\begin{equation}
\Xi =\sum_{n}\Xi_{n}.  \label{sum}
\end{equation}%
This sum has $37^{N}$ elements. One can easily find the expected return of
the roulette, $\Omega$, as 
\begin{equation}
\Omega =36\sum_{n}\frac{\Xi _{n}}{j_{n}}-1.
\end{equation}

\section*{Results and discussions}

In order to analyze the chances of a gambler to have an advantage over
casino, let us start from the critical spread of the probability
distribution $\delta$ which equalises the chances of casino and gambler. It
is given by a condition: 
\begin{equation}
\Omega = 0.
\end{equation}

An important particular case is the limit of an ideal roulette. In this case 
$\delta = 0$, $P(k) =1/37$ for all $k$, $\Xi_{n}=\frac{j_{n}}{37^{N+1}}$.
Clearly, in this case the casino has an advantage over the gambler as 
\begin{equation}
\sum_{n}\Xi_{n} = \frac{1}{37^{N}}\sum_{n}\frac{j_{n}}{37} < \frac{1}{37^{N}}%
\sum_{n}\frac{j_{n}}{36}.
\end{equation}
In this limit, the expected return of the roulette $\Omega_{0}$\ is well
known: 
\begin{equation}
\Omega_{0}=\frac{36}{37} - 1 \approx -0.02703.
\end{equation}

Another limit is of $\delta =\infty$ that corresponds to a zero probability
of realisation of all numbers with $k<18$, $P(18)=1/37$ and $P(k>18)=2/37$.
In this case the gambler always wins with the adapted method.

In Fig. \ref{Fig1}(c,d), we show the results of Monte-Carlo simulations\ 
\cite{MCM} of the expected return of the roulette $\Omega(\delta)$ and $%
\Omega(\xi)$, respectively, calculated for different values of $N$. We have
used a standard generator of a sequence of uniformly distributed random
numbers~\cite{MKL}, and run $10^{6}$ realisations for each point. One can
see that starting from $\delta \approx 0.05$ a gambler has an advantage over
the casino for a wide range of $N$. This critical value of $\delta$
corresponds to the ratio of probabilities of the realisation of the highest
probable and the lowest probable numbers of $\xi_{c}\approx 2.012$.

\begin{figure}[tbp]
\includegraphics[width=.6\columnwidth]{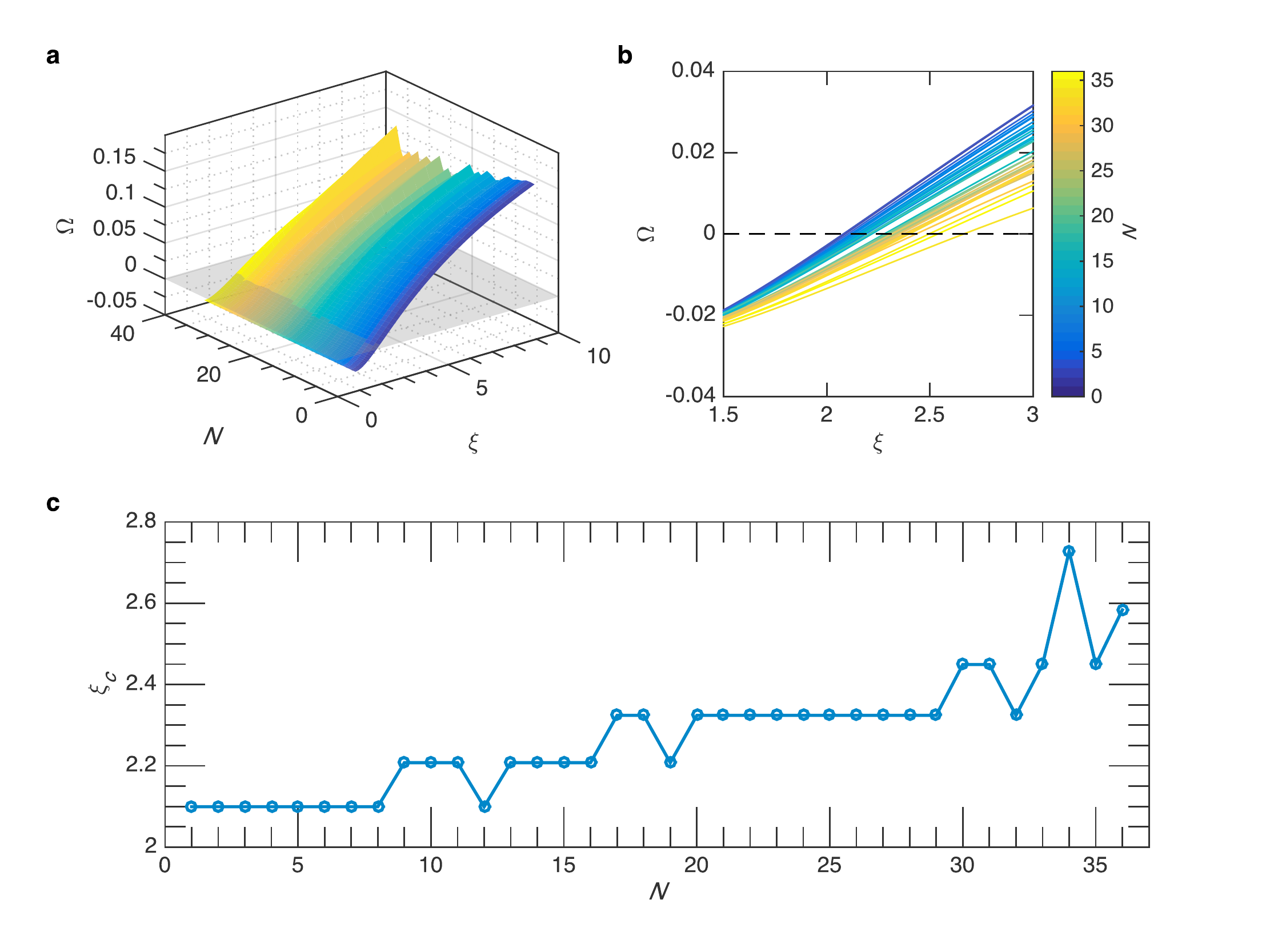}
\caption{The Monte-Carlo simulation of the expected return $\Omega$ of a
non-ideal roulette versus (a) the length of the sequence of the "last"
numbers $N$ used by the gambler to place bets, and the lowest-to-highest
probability ratio, $\protect\xi$ (b) shows the same as a 2D plot for the
convenience. (c) Monte-Carlo simulation of the critical ratio of
probabilities of realsations of the most probable and least probable number, 
$\protect\xi_{c}$, as a function of $N$. }
\label{Fig2}
\end{figure}

Fig. \ref{Fig2}(a) is a 3D plot showing $\Omega (N, \xi)$. From the 2D plot
of the same function shown in Fig. \ref{Fig2}(b) one can see that, in
general, $\Omega $ slightly decreases with the increase of $N$, and a
gambler can have an advantage over the casino for a wide range of $N$ if $%
\delta > 0.05$ ($\xi > 2$). Fig. \ref{Fig2}(c) shows $\xi_{c}$ that provides
the expected return $\Omega =0$ as a function of $N$.

\begin{figure}[tbp]
\includegraphics[width=.6\columnwidth]{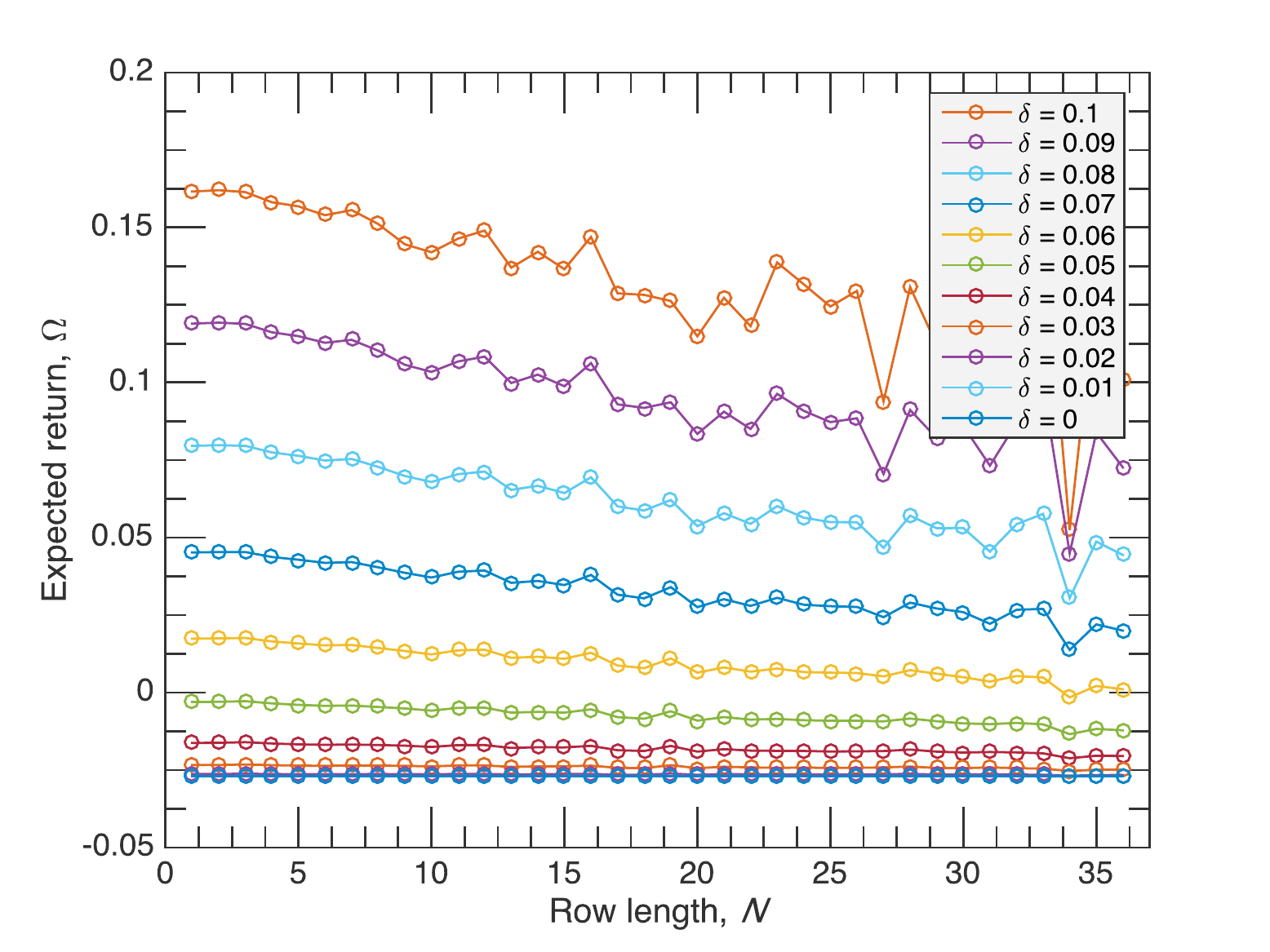}
\caption{Monte-Carlo simulation of the expected return $\Omega$ versus the
length of the sequence of the ``last'' numbers $N$ for different values of $%
\protect\delta$ and lowest-to-highest probability ratio, $\protect\xi$.}
\label{Fig3}
\end{figure}

The dependence of $\Omega $ on $N$ is important and needs further analysis.\
In Fig. \ref{Fig3}, we show the results of Monte-Carlo simulations of the
expected return $\Omega$ as a function of $N$ for different values of the
probability spread $\delta$. One can see that for small $\delta$ the
expected return does not show any appreciable dependence on $N$, while for
larger $\delta$ the expected return slightly decreases with the increase of $%
N$. It looks like the choice of a relatively small $N$ is the best. However,
as we will below, small $N$ strategies are more risky in terms of a
probability of critical fluctuations that would consume the entire capital
of the gambler. 

\subsection*{Linear distribution function}

In order to check the robustness of our method, we have also run the
simulations for the expected return using a simplest linear probability
distribution function: 
\begin{equation}
P\left( k\right) =\frac{1}{37}\left( 1+\frac{k-18}{18}\beta \right) ,
\label{linear}
\end{equation}%
where $\beta$ is a parameter describing the spread of the probability. 
\begin{figure}[tbp]
\includegraphics[width=.7\columnwidth]{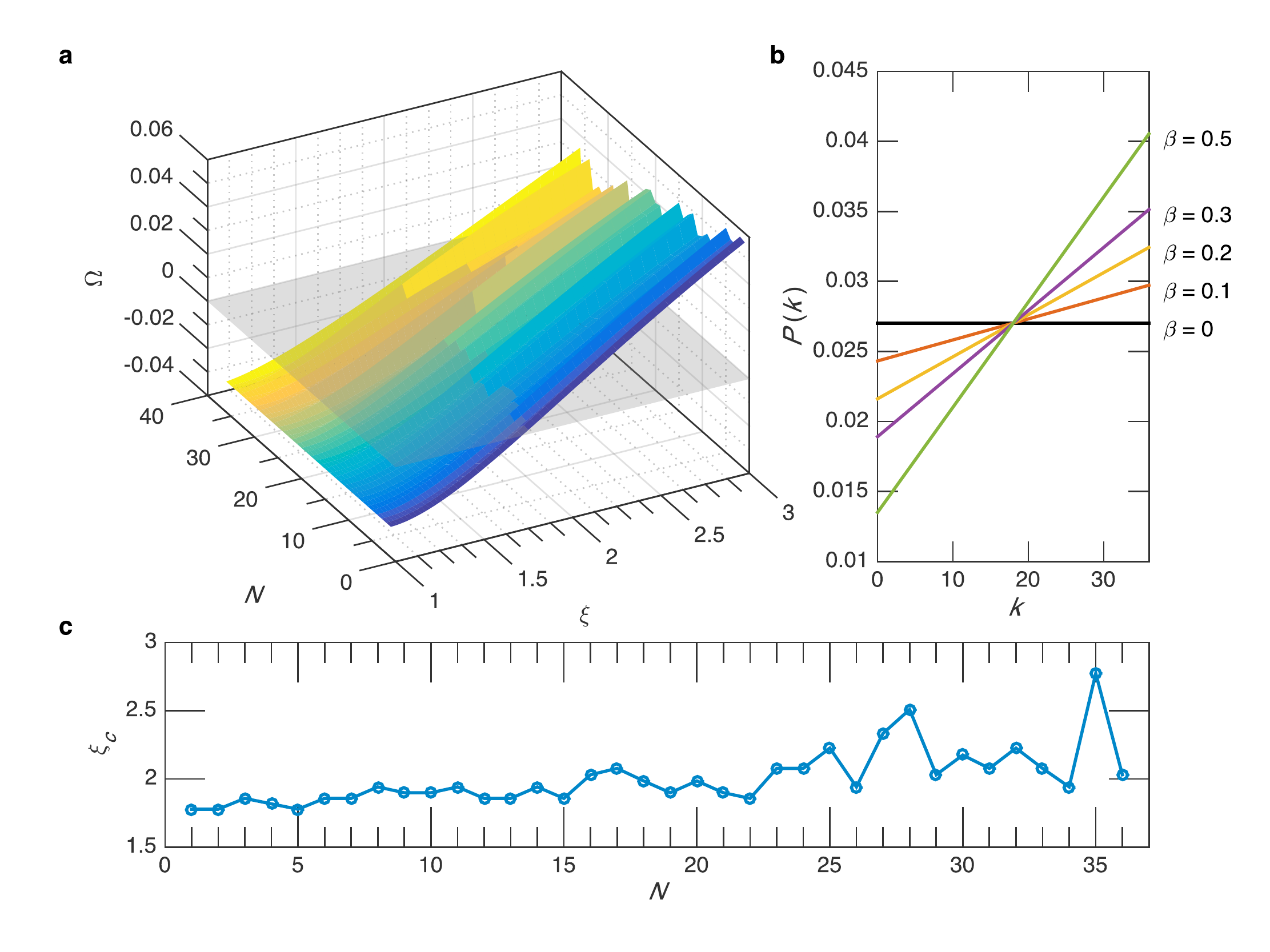}
\caption{(a)~Monte-Carlo simulation of the expected return $\Omega$ as a
function of the length of the row of ``last'' numbers $N$ used by a gambler
when placing bets and of the probability ratio $\protect\xi$. The gray plane
indicates $\Omega = 0$. The calculation has been done using a linear
probability distribution $P(k)$ given by Eq.~\eqref{linear}. (b) shows the
functions $P(k)$ taken at different values of the spread parameter $\protect%
\beta$. (c)~The critical value of highest-to-lowest probability ratio, $%
\protect\xi_{c}$, versus $N$. }
\label{Fig4}
\end{figure}
In this case, 
\begin{equation}
\xi =\frac{P\left( 36\right) }{P\left( 0\right) }=\frac{1+\beta }{1-\beta }.
\end{equation}

Fig. \ref{Fig4}(a) shows the 3D plot $\Omega (N,\xi)$ calculated using the
probability function \eqref{linear}. Qualitatively, the results of the
modeling are similar to those obtained with the Gaussian distribution.
Indeed, the spread of probabilities results in bunching of numbers, so that
a gambler playing on the last $N$ numbers may have an advantage over casino.
The critical value of $\xi$ that corresponds to $\Omega=0$ is close to 2
similarly to the results obtained with the distribution function~%
\eqref{eq_Gauss_distribution}. The dependence of $\Omega$ on $N$ is very
weak, within the statistical noise produced by our Monte-Carlo simulator.

Finally, note that our analysis can be easily generalised for the case of an
American roulette by replacing 37 by 38 in the above formalism. Obviously,
the critical value of $\delta$ would be higher for the American roulette as
compared to the European roulette.


\subsection*{Critical initial capital}

For the practical implementation of the strategy described here, it is
important to know the value of an initial capital that assures successful
gambling. We recall that in the case of an ideal roulette that is
characterised by a negative $\Omega \approx -0.027$ for any finite starting
capital the gambler loses after a sufficiently large number of bets. In our
case, $\Omega >0$, and the capital of a gambler normally increases in the
course of the game. However, the risk is also present: due to a negative
fluctuation (sequence of lost spins) the gambler can lose his/her capital
and be put out of a game. What should be the size of the initial capital
that ensures the relatively safe game and reduces the risk of catastrophic
negative fluctuations? While it is impossible to exclude a catastrophic
fluctuation entirely, one can introduce such a characteristic (critical)
value of the initial capital $C$. We define $C$ as a capital that allows a
gambler to double it during the average number of spins $M$ that elapses
between two catastrophic fluctuations. A catastrophic fluctuation is defined
as a loss of $C$ in a sequence of unsuccessful spins. We note at this point
that for an ideal roulette the critical capital does not exist: the
probability to lose any capital is always higher than the probability to
double it by at least 1/37. One can easily see that $C$ is linked with $M$ by a
simple relation:%
\begin{equation}
C=M\overline{\jmath }(N)\Omega ,  \label{capital}
\end{equation}%
where $\overline{\jmath }(N)$ is the average value of different numbers
among the last $N$ numbers. Now, the frequency of catastrophic fluctuations
is given by%
\begin{equation}
f=\left( \frac{37-\overline{\jmath }(N)(1+\Omega )}{37}\right) ^{S},
\end{equation}%
where $S$ is the number of successive unsuccessful spins needed to lose the
capital $C$. As each next unsuccessful spin increases our loss by $\overline{%
\jmath }(N)$, in average, we obtain:%
\begin{equation}
C=S\overline{\jmath }(N),  \label{capital1}
\end{equation}

Now, $M$ is linked with the frequency of catastrophic events by 
\begin{equation}
M=1/f.
\end{equation}
Expressing $M$ and $S$ in terms of $C$ with use of Eqs.\ \eqref{capital} and %
\eqref{capital1}, we obtain the transcendental equation for the critical
capital:

\begin{equation}
\frac{\overline{\jmath}(N)\Omega}{C}=\left(\frac{37-\overline{\jmath}%
(N)(1+\Omega)}{37}\right)^{C/\overline{\jmath}(N)}.
\end{equation}%
Here, $C$ is expressed in units of bets. To be on a safe side, a gambler
should possess a capital exceeding $C$. 
\begin{figure}[tbp]
\includegraphics[width=.95\columnwidth]{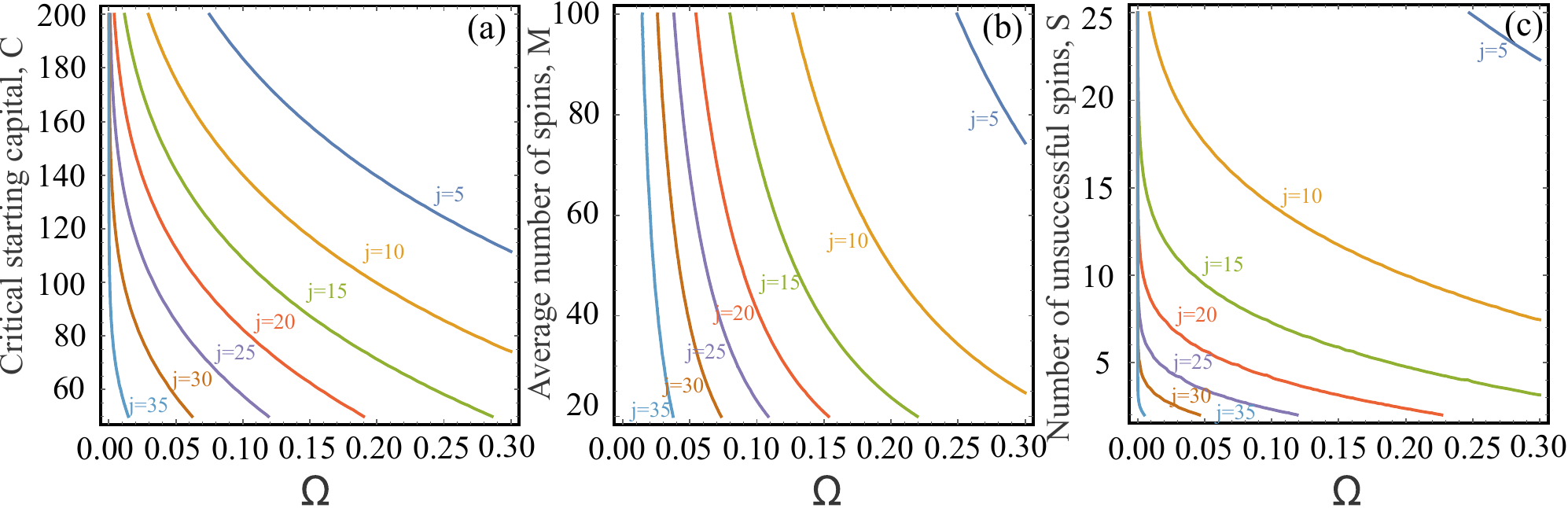}
\caption{The critical starting capital $C$ (a), the corresponding average
number of spins $M$ between two catastrophic fluctuations (b), and number of
successive unsuccessful spins $S$ needed to spend the critical capital (c)
as functions of $\Omega$ calculated for the different values of the average
number of bets per spin $\overline{\jmath}(N)$. This calculation has been
done with the probability distribution function~\eqref{eq_Gauss_distribution}%
.}
\label{Fig5}
\end{figure}

Figure \ref{Fig5} shows the dependences of $C$ and corresponding values of $M
$ and $S$ on $\Omega$ calculated for different values of the average number
of bets per spin $\overline{\jmath}(N)$. One can see that $C$ strongly
decreases with the increase of both $\Omega$ and $\overline{\jmath}(N)$.
This is important for a gambler who aims at following the safetest strategy,
while willing to achieve significant benefits nevertheless. One should keep
in mind also that $\Omega$ slightly decreases as a function of $N$, while $%
\overline{\jmath}(N)$ is a monotonically increasing function. Clearly,
playing on a larger series of latest numbers $N$ one can strongly reduce the
risk of a catastrophic fluctuation. The price to pay is a simultaneous
decrease of the expected return. The most convenient range of $N$ is likely
to be between 10 and 20 for a sufficiently high $\delta$ ($\delta$ is of the
order of $0.1$).

Note also that the avarage number of spins separating two negative
fluctuations which lead to the loss of the initial capital, $M$, as well as
the number of successive unsucessful spins needed to exhaust this capital, $S
$,\ linearly increase with the increase of $C$, as Figure 5(b,c) shows. Both
quantities decrease with the increase of the expected return $\Omega$ and
with the increase of the average number of bets per spin $\overline{\jmath}%
(N)$.

\subsection*{Practical tests}

The strategy described here has been tested on three European roulette
tables in two different casinos of Southampton (UK) on several different
days. We have played $N=12$ in one casino and $N=17$ in another one. In both
cases the results were steadily positive with $\Omega \approx 0.1$
independently on croupier. The statistics has been taken over several
hundred of spins. For evident reasons, we did not inform administration of
the casinos of our experiments, which could have been run without any
external bias. The tests have been realised with minimum allowed bets and
all the money won in this way were donated to charity.

\section*{Conclusions}

To conclude, our analysis shows that certain strategies may result in the
positive expected return in the case of a non-ideal roulette. Our method of
playing on the last $N$ numbers does not require any preliminary knowledge
on \textquotedblleft hot\textquotedblright\ or \textquotedblleft
cold\textquotedblright\ numbers, speed of the ball, speed of the wheel etc.
The outcome of a long game is only dependent on the value of the probability
spread parameters $\delta $ or $\xi $\ and on the chosen value of $N$. We
show also that while the expected return of a non-ideal roulette shows only
a weak dependence on $N$ (it slightly decreases with the increase of $N$), a
gambler may significantly reduce the risk of a catastrophic loss if playing
on a longer sequence of numbers. In practical terms, we would recommend a
gambler to play the strategy described here with any unknown roulette by
initially placing only minimum bets for the first 100 or 200 spins (the
higher $N$ is chosen the lower number of spins is needed). If the outcome
turns out to be steadily positive, it means that, most likely, $\delta $ is
above critical. Once all doubts are left behind, the gambler may start
placing maximum bets that would bring him a significant benefit. If, when
placing minimum bets for 100 or 200 spins the gambler does not see any sign
of a steadily positive tendency, the chosen roulette is probably close to
ideal, and there is no reason to continue playing on this roulette. We also
warn that the positive expected return does not guarantee a gambler against
the loss due to a negative fluctuation and advice against irresposible
gambling.

\acknowledgments

ASS acknowledges ITMO Fellowship and Visiting Professorship.

\end{document}